\documentclass[runningheads]{llncs}
\usepackage{marvosym}
\usepackage{multirow}
\usepackage{graphicx}
\usepackage{wrapfig}
\usepackage{amssymb}
\usepackage{wrapfig}
\begin{document}

\title{EnMcGAN: Adversarial Ensemble Learning for 3D Complete Renal Structures Segmentation}
\titlerunning{3D Complete Renal Structures Segmentation}
\author{Yuting He\inst{1} \and Rongjun Ge\inst{1} \and Xiaoming Qi\inst{1} \and Guanyu Yang\inst{1,3}(\Letter) \and Yang Chen\inst{1,3} \and Youyong Kong\inst{1,3} \and Huazhong Shu\inst{1,3} \and Jean-Louis Coatrieux\inst{2} \and Shuo Li\inst{4}}

\authorrunning{Y. He et al.}
\institute{LIST, Key Laboratory of Computer Network and Information Integration (Southeast University), Ministry of Education, Nanjing, China
\email{yang.list@seu.edu.cn}\and
Univ Rennes, Inserm, LTSI - UMR1099, Rennes, F-35000, France \and
Centre de Recherche en Information Biomédicale Sino-Français (CRIBs) \and
Dept. of Medical Biophysics, University of Western Ontario, London, ON, Canada}
\maketitle              % typeset the header of the contribution
\begin{abstract}
3D complete renal structures(CRS) segmentation targets on segmenting the kidneys, tumors, renal arteries and veins in one inference. Once successful, it will provide preoperative plans and intraoperative guidance for laparoscopic partial nephrectomy(LPN), playing a key role in the renal cancer treatment. However, no success has been reported in 3D CRS segmentation due to the complex shapes of renal structures, low contrast and large anatomical variation. In this study, we utilize the adversarial ensemble learning and propose \emph{Ensemble Multi-condition GAN}(EnMcGAN) for 3D CRS segmentation for the first time. Its contribution is three-fold. \textbf{1)}Inspired by windowing\cite{goldman2007principles}, we propose the multi-windowing committee which divides CTA image into multiple narrow windows with different window centers and widths enhancing the contrast for salient boundaries and soft tissues. And then, it builds an ensemble segmentation model on these narrow windows to fuse the segmentation superiorities and improve whole segmentation quality. \textbf{2)}We propose the multi-condition GAN which equips the segmentation model with multiple discriminators to encourage the segmented structures meeting their real shape conditions, thus improving the shape feature extraction ability. \textbf{3)}We propose the adversarial weighted ensemble module which uses the trained discriminators to evaluate the quality of segmented structures, and normalizes these evaluation scores for the ensemble weights directed at the input image, thus enhancing the ensemble results. 122 patients are enrolled in this study and the mean Dice coefficient of the renal structures achieves 84.6\%. Extensive experiments with promising results on renal structures reveal powerful segmentation accuracy and great clinical significance in renal cancer treatment.
\end{abstract} 

\section{Introduction}
\begin{figure*}[htb]
\centering
\includegraphics[width=10cm]{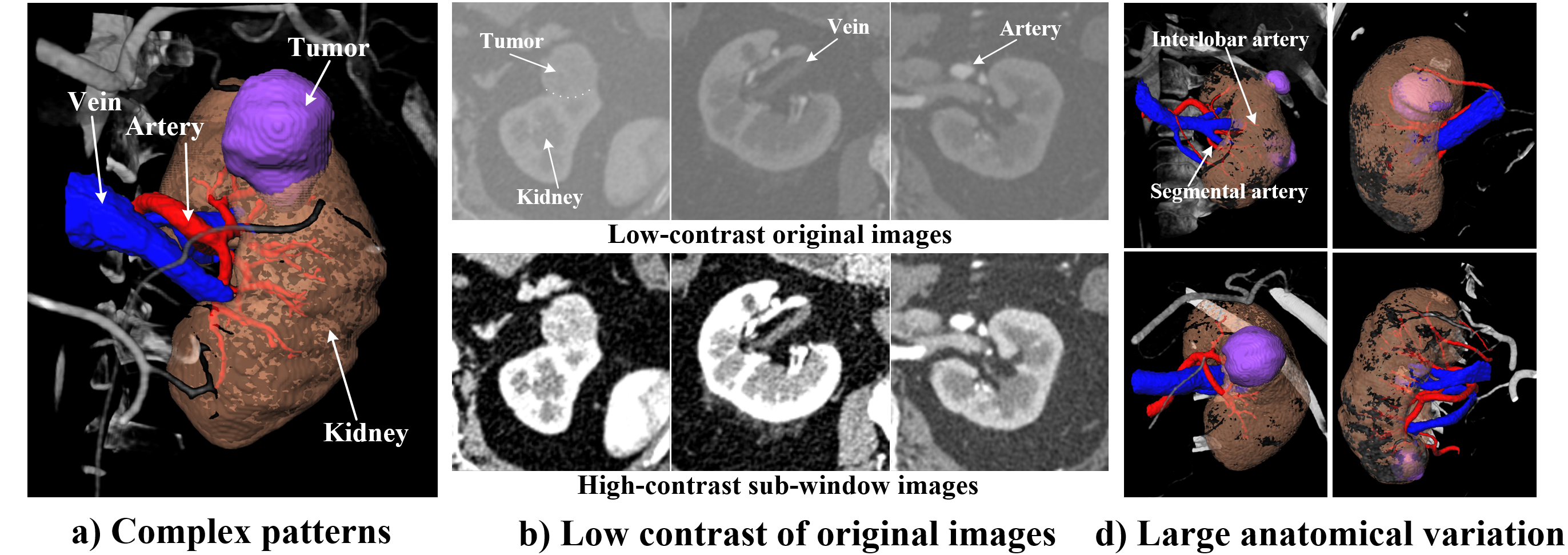}
\caption{The challenges of the 3D CRS segmentation. a) The renal structures has complex shapes making difficult feature extraction. b) The original CTA images has low contrast bringing low segmentation quality on boundaries. c) The renal structures has large anatomical variation in different cases causing weak generalization ability.}
\label{fig1}
\end{figure*}
3D complete renal structures(CRS) segmentation on CTA image targets on segmenting the kidneys, tumors, renal arteries and veins in one inference. Once successful, it will provide preoperative plans and intraoperative guidance for laparoscopic partial nephrectomy (LPN) \cite{Shao2011Laparoscopic,shao2012precise}, playing a key role in renal cancer treatment \cite{porpiglia2018hyperaccuracy}. \emph{Preoperatively}, the renal artery reaching the interlobar arteries will guide the estimation of perfusion regions to select the tumor-feeding branches and locate the arterial clamping position \cite{Shao2011Laparoscopic}. The kidney and tumor will show the lesions' location to pre-plan the tumor resection surface \cite{zhang2019application}. \emph{Intraoperatively}, the veins outside the hilum will help the clinicians exclude the unconcerned vessels thus clamping the accurate arterial positions quickly. Besides, the 3D CRS visual model will also be fused with the laparoscopic videos bringing the augmented reality \cite{nicolau2011augmented}, so that the invisible regions will be supplemented to guide the operation smoothly. With the assistance of 3D CRS segmentation, the safety of renal surgery is improved, the pain of patients is relieved and the cost of the treatment is reduced.

However, there is no effective solution for 3D CRS segmentation. Taha et al \cite{Taha2018Kid} utilized a Kid-Net achieving the renal vessels segmentation. He et al \cite{he2019dpa} proposed a semi-supervised framework and achieved the fine renal artery segmentation. Li et al \cite{Li2018Segmentation} designed a Residual U-Net for renal structures segmentation. The related works are limited in our task: \textbf{1)}On one hand, some works \cite{Taha2018Kid,he2019dpa} only focused on partial renal structures lacking the countermeasures to the complete structures in one-inference segmentation. The one-by-one segmentation also will make overlap of different structures especially on low-contrast boundaries. \textbf{2)}On the other, some works \cite{Li2018Segmentation,Taha2018Kid} lack the fine detail of the segmented structures limiting the clinical downstream tasks. For example, their arteries only reaches the segmental arteries losing the ability of perfusion regions estimation.

Formidable challenges of 3D CRS segmentation are limiting its further applications: \textbf{1)} Complex shapes of renal structures. These structures have complex shapes, for example, the arteries have tree-like shape while the tumors have ball-like shape (Fig.~\ref{fig1} (a)). The model has to represent their features simultaneously resulting in a difficult feature extraction process and limiting its generalization ability. \textbf{2)} Coarse-grained pattern and low contrast. CT image has a large gray range (4096), and the structures of interests in our task are only in a narrow range. This makes the contrast of soft tissues and the boundary regions are low in original images (Fig.~\ref{fig1}(b)) such as the tumors and veins, making coarse-grained pattern. Therefore, the segmentation network will difficult to perceive the fine-grained pattern in such a narrow distribution limiting the integrity of the structures and the quality of boundaries. \textbf{3)} Large anatomical variation. Renal vessels have uncertain branch numbers and growth topology \cite{he2019dpa,Petru2013Erratum}, and the location of the tumors and their damage to the kidneys are uncertain(Fig.~\ref{fig1}(c)). These anatomical variation makes it difficult to cover all variation, limiting the model's generalization.

The windowing \cite{goldman2007principles} removes irrelevant gray ranges and expands the interested distribution making the network focus on the wider distribution to perceive fine-grained pattern. Inspired by the radiologist delineating the renal structures in different window widths and centers (sub-windows) via windowing, we propose the \emph{multi-windowing committee (MWC)}. We select the sub-windows which of superiority in renal structures on CTA images, and divide the CTA image into these sub-windows thus expanding our task-interested distribution and making fine-grained pattern such as boundaries (Fig.~\ref{fig2}(a)). We train multiple learners on these sub-window images for the fine-grained representations on their covered distributions making segmentation superiorities. Finally, these superiorities will be fused \cite{polikar2012ensemble} improving the integrated segmentation quality.

The shape regularisation utilizes the shape knowledge learned by an additional model to encourage the CNN extracting higher-order shape features \cite{mosinska2018beyond,oktay2017anatomically,luc2016semantic}. We propose the \emph{Multi-condition GAN (McGAN)} which equips the segmentation model with multiple discriminators for global shape constraints (Fig.~\ref{fig2}(b)). During adversarial training, the discriminators learn to evaluate the similarity between the segmented structures' shapes and real shapes, thus learning the higher-order shape knowledge of the renal structures via the min-max game of the adversarial training \cite{goodfellow2014generative}. Then, these knowledge are made as the conditions that the segmented renal structures have to meet when optimizing the segmentation models, improving the shape feature extraction ability.

Fusing the segmentation superiorities of the learners will bridge their representation preferences caused by large anatomical variations \cite{polikar2012ensemble,dietterich2002ensemble}. These preferences will make the learners have better segmentation quality in their advantageous regions and poor quality in vulnerable regions, if being fused, the whole accuracy will be improved. Therefore, we propose the \emph{adversarial weighted Ensemble (AWE)} (Fig.~\ref{fig2}(c)) which utilizes the trained discriminators to evaluate the pixel-wise segmentation quality of the results, thus giving higher weights to better quality regions and making the ensemble direct at the segmentation quality dynamically. So, the fusion of the advantageous regions in the results will balance the representation preferences and bring personalized ensemble process.

We propose an adversarial ensemble learning framework, EnMcGAN, for 3D CRS segmentation for the first time. Our detailed contributions are as follow:
\begin{itemize}
\item We propose the adversarial ensemble learning which equips the ensemble segmentation model with adversarial learning for the first time, and propose the \emph{EnMcGAN} for 3D CRS segmentation which will play an important role in accurate preoperative planning and intraoperative guidance of LPN. Our complete experiments demonstrate its excellent performance.

\item We propose the \emph{multi-windowing committee (MWC)} which divides CTA image into the sub-windows which of superiority in renal structures,  making fine-grained pattern. Then the learners trained on these sub-windows are constructed as an ensemble framework to fuse the fine-grained representations on the covered distributions of these sub-windows, thus improving integrated segmentation quality.

\item We present the \emph{multi-condition GAN (McGAN)} which embeds the shape knowledge in segmentation model via multiple discriminators, thus encouraging the segmented results being consistent with their shape prior knowledge and improving the shape features extraction ability.

\item We propose the \emph{adversarial weighted Ensemble (AWE)} which utilizes segmentation quality evaluation ability of the trained discriminators to fuse the advantageous regions in the segmented results, thus bringing personalized fine ensemble process and balancing the representation preferences.
\end{itemize} 

\section{Methodology}
\label{Methodology}
\begin{figure*}[htb]
\centering
\includegraphics[width=10cm]{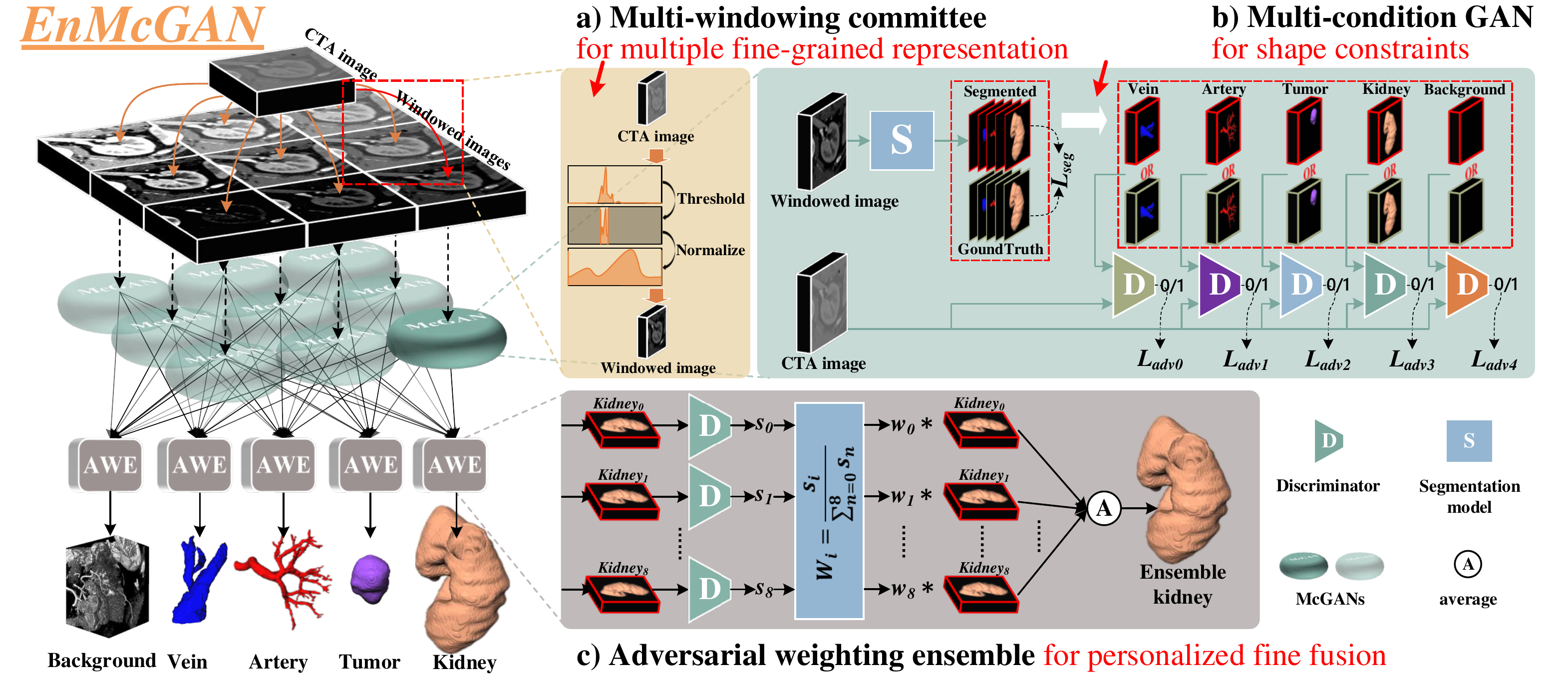}
\caption{The illustration of our EnMcGAN: a) MWC enhances the contrast on windowed images bringing salient boundaries and soft tissues (Sec.~\ref{sec1}); b) McGAN utilizes multiple discriminators for shape constraints (Sec.~\ref{sec2}) and c) AWE module utilizes the trained discriminators for dynamic ensemble weights (Sec.~\ref{sec3}).}
\label{fig2}
\end{figure*}
As shown in Fig.~\ref{fig2}, our proposed EnMcGAN takes adversarial ensemble learning for high quality 3D CRS segmentation. It has three cooperative elements: \textbf{1)} Multi-windowing committee (MWC, Sec.~\ref{sec1}) thresholds and normalizes the CTA image enhancing the contrast and making the fine-grained pattern, and fuses the segmentation superiorities of multiple segmentation learners trained in different covered distributions improving the integrated segmentation quality. \textbf{2)} Multi-condition GAN (McGAN, Sec.~\ref{sec2}) utilizes shape constraints from the discriminator to encourage the segmented renal structures meeting their real shape conditions, thus improving the shape features extraction ability. \textbf{3)} Adversarial weighted Ensemble (AWE, Sec.~\ref{sec3}) utilizes the trained discriminator for dynamic ensemble weights, thus and balancing the representation preferences and providing the personalized ensemble results.
\subsection{Multi-windowing committee for multiple fine-grained representation}
\label{sec1}
Our MWC (Fig.~\ref{fig2}(a)) divides CTA image to multiple narrow windows \cite{goldman2007principles} and expends their distributions making the significant region and fine-grained pattern. And then it fuse the segmentation superiorities from the distributions covered by multiple sub-windows, thus improving integrated segmentation quality.

\textbf{Windowing for fine-grained pattern.} The CTA image $x$ is divided to multiple narrow windows with different window centers and widths. Firstly, nine different windows are selected. We segment CTA images via k-means clustering \cite{ng2006medical} for $c$ categories, in our experiment $c=5$. Then, the mean CT values of these categories in our dataset are calculated and selected for three window centers $c$, in our experiment, $\{c_{0}=1032.7834,c_{1}=1150.0825,c_{2}=1332.1959\}$. We also define three default window widths $\{w_{0}=256,w_{1}=512,w_{2}=768\}$, thus combining for nine sub-windows. Then, the CTA image $x$ is thresholded to remove irrelevant gray range and normalized to expand the interested distribution via windowing, thus making nine windowed images $\{x_{0},x_{1},...,x_{8}\}$: $x_{i}=\frac{max(min(x,c_{j}-\frac{w_{k}}{2}),c_{j}+\frac{w_{k}}{2})-(c_{j}-\frac{w_{k}}{2})}{w_{k}}$. Finally, these images are used to train nine segmentation learners $S(\dot)$ which will learn the fine-grained representations on the distributions covered by these sub-windows.

\textbf{Summary of the advantages.} \textbf{1)} The CTA image is thresholded with different window centers and widths for small gray ranges and normalized to $[0,1]$, thus the irrelevant gray range will be removed and the interested distribution will be expended. Therefore, these sub-windows will have significant regions and fine-grained patterns improving the segmentation quality. \textbf{2)} The image in different narrow windows will have different significant regions, so our multiple learners trained on these sub-window images will learn the fine-grained representations on their covered distributions making segmentation superiorities. These superiorities are fused improving the final segmentation quality.
\subsection{Multi-condition GAN for shape constraints}
\label{sec2}
Our McGAN (Fig.~\ref{fig2}(b)) utilizes multiple discriminators to provide the segmentation model shape constraints, thus encouraging the segmented structures meeting their real shape conditions, improving the shape feature extraction ability.

\textbf{McGAN for adversarial segmentation.} In training stage, our McGAN inputs the real or segmented renal structures together with the CTA image into five discriminators resulting in a conditional GAN \cite{mirza2014conditional} for shape constraints of renal structures. The segmentation model $S$ takes the DenseBiasNet \cite{he2019dpa} which fuses the multi-receptive field features for the multi-scale feature representation ability. The discriminators $\{D_{0},...,D_{4}\}$ follow the 3D version of the VGG-A \cite{simonyan2014very} and activated by sigmoid function. As shown in Fig.~\ref{fig2} (b), it takes the original image $x$ and the segmented $S(x_{i})_{n}$ or ground truth $y_{n}$ mask of each structure are input to discriminator to learn the shape constraint of each renal structure and the evaluation ability of segmentation quality. The binary cross-entropy loss $\mathcal{L}_{bce}$ is calculated as the adversarial loss $\mathcal{L}_{adv_n}$ of each structure:
\begin{equation}\label{equ1}
  \mathcal{L}_{adv_n}(\theta_{D_{n}})=\mathcal{L}_{bce}(D_{n}(x,y_{n}),1)+\mathcal{L}_{bce}(D_{n}(x,S(x_{i})_{n}),0).
\end{equation}
The segmentation model takes the windowed images $x_{i}$ as input and is optimized by the multi-class cross-entropy loss $\mathcal{L}_{mce}$ from the segmented results $S(x)$ and labels $y$, together with the adversarial losses corresponding to renal structures from the discriminator. Therefore, the hybrid segmentation loss is:
\begin{equation}\label{equ2}
  \mathcal{L}_{seg}(\theta_{S})=\mathcal{L}_{mce}(S(x_{i}), y)+\lambda\sum^{N}_{n=0}\mathcal{L}_{bce}(D_{n}(x,S(x_{i})_{n}),1),
\end{equation}
where the $\lambda$ is the weight to balance loss functions in this hybrid loss. In our experiment, it is 0.01. The $\theta_{S}$ and $\theta_{D_{n}}$ is the parameters of the segmentation model and discriminators. As an ensemble model, narrow-window images are used to train nine McGANs which share their discriminators iteratively.

\textbf{Summary of the advantages.} During adversarial training, the discriminators learn to evaluate the similarity between the segmented structures' shapes and real shapes. When optimizing the segmentation model, the real shapes will become the conditions encoded by the discriminators, so the adversarial loss will encourage the segmented structures to meet them improving the shape feature extraction ability of the segmentation model.
\subsection{Adversarial weighted ensemble for personalized fine fusion}
\label{sec3}
Our AWE module (Fig.~\ref{fig2}(c)) utilizes the trained discriminators from our McGAN to evaluate the segmentation quality of renal structures and generate the dynamic ensemble weights directed at the input image, thus bringing personalized fine ensemble process and balancing the representation preferences.

\textbf{Ensemble process of AWE.} In testing stage, our AWE module fuses the results based on their segmentation quality, improving the ensemble results. Fig.~\ref{fig2} (c) illustrates the ensemble process of this module taking the kidney as an example. The segmented kidneys $\{kidney_{0},...,kidney_{8}\}$ from the segmentation learners together with the original image $x$ are putted into the trained discriminators for evaluation scores $\{s_{0},..,s_{8}\}$. Then, these scores are normalized via $w_{i}=\frac{s_{i}}{\sum^{8}_{n=0}s_{n}}$ for the ensemble weights directed at the input image. Finally, these weights are used to weight the average ensemble enhancing the results: $\hat{y_{kidney}}=\frac{1}{9} \sum_{i=0}^{9} kidney_{i}*w_{i}$.

\textbf{Summary of the advantages.} The trained discriminators evaluate the segmentation qualities of renal structures, so we utilize them for dynamic ensemble weights directed at input images. Therefore, the incorrect weights are avoided, and the ensemble results are effectively enhanced.

\section{Materials and Configurations.}
\label{Materials}
\textbf{Dataset.} 122 patients with renal cancer treated with LPN are enrolled in this study. The kidney region of interests (ROIs) with tumors on their CTA images were extracted as the dataset in our experiments. Their pixel sizes are between $0.47mm$ and $0.74mm$, the slice thickness is $0.5mm$ and the image size is $150\times150\times200$. Five kidney tumor subtypes including clear renal cell carcinomas, papillary, chromophobe, angiomyolipoma and eosinophilic adenoma are included in this dataset resulting in large heterogeneity and anatomical variation, and the tumor volume is vary up to 300 times. The kidney, tumor, vein and artery on these images are fine labeled.

\textbf{Comparison settings.} To evaluate the superiority of our framework, we perform extensive experiments. The V-Net \cite{milletari2016v}, 3D U-Net \cite{o20163D}, Res-U-Net \cite{Li2018Segmentation}, Kid-Net \cite{Taha2018Kid}, DenseBiasNet \cite{he2019dpa} and the ensemble model, VFN \cite{xia2018bridging}, are trained as the comparison models. For fair comparison, VFN also takes DenseBiasNet as the segmentation learners. The Dice coefficient (DSC) $[\%]$, mean surface voxel distance (MSD) are used to evaluate the coincidence of the region and surface. Besides, the mean centerline voxel distance (MCD) is used to evaluate the coincidence of the artery topology following \cite{he2020dense}.

\textbf{Implementation.} During training, $150\times150\times128$ patches are cropped in the z-axis so that the dataset is enlarged and the GPU memory is saved. Each model is optimized by Adam \cite{kingma2015adam} with the batch size of 1, learning rate of $1\times10^{-4}$ and iterations of 40000. The 5-fold cross-validation is performed for comprehensive evaluation. All methods are implemented with PyTorch and trained on NVIDIA TITAN Xp GPUs.
\section{Results and analysis}
\label{Results}
Our EnMcGAN brings the fine-grained pattern and significant regions, embeds shape priori knowledge into segmentation model and bridge the representation preferences caused by large anatomical variations thus achieving the excellent 3D complete renal structures segmentation. In this part, we will thoroughly evaluate and analyze the effectiveness of our proposed EnMcGAN: \textbf{1)} The quantitative evaluation, qualitative evaluation and the ROC and PR curves in comparative study (Sec.~\ref{Comparative}) will show the superiorities of our framework compared with other models. \textbf{2)} The ablation study (Sec.~\ref{Ablation}) will demonstrate the contribution of each our innovation in our framework. \textbf{3)} The the performances of each segmentation learner and the number of fused learners will be analysed in framework analysis (Sec.~\ref{Framework}).
\subsection{Comparative study shows superiority}
\label{Comparative}
\begin{table*}[htb]
\centering
\caption{Evaluation on renal structures reveal powerful performance of our framework. The 'En' means the method is the ensemble learning model.} %
\resizebox{\textwidth}{!}{
\begin{tabular}{l|cc|cc|cc|ccc|c}
\hline
\multirow{2}{*}{Method}             &\multicolumn{2}{c|}{Kidney}&\multicolumn{2}{c|}{Tumor}&\multicolumn{2}{c|}{Vein}&\multicolumn{3}{c|}{Artery}&Mean DSC\\
\cline{2-10}
                                    &DSC&MSD&DSC&MSD&DSC&MSD&DSC&MCD&MSD&$\pm$std\\
\hline
V-Net \cite{milletari2016v}          &94.1$\pm$2.3&1.17$\pm$0.72&67.5$\pm$26.8&7.01$\pm$6.36&66.9$\pm$14.8&4.57$\pm$6.35&83.0$\pm$8.2&3.21$\pm$2.66&2.03$\pm$2.81&77.9$\pm$7.8\\
3D U-Net \cite{o20163D}              &91.9$\pm$10.9&1.11$\pm$0.60&72.1$\pm$26.3&5.22$\pm$6.12&65.4$\pm$20.8&2.41$\pm$1.60&80.5$\pm$9.9&1.93$\pm$1.06$\pm$&1.30$\pm$1.28&77.5$\pm$12.2\\
Res-U-Net \cite{Li2018Segmentation}  &92.5$\pm$3.8&1.63$\pm$0.95&51.2$\pm$29.7&13.02$\pm$16.73&63.4$\pm$17.7&3.00$\pm$1.73&81.9$\pm$6.1&2.95$\pm$2.00&1.91$\pm$2.00&72.2$\pm$7.9\\
Kid-Net \cite{Taha2018Kid}           &91.0$\pm$11.5&1.49$\pm$1.14&69.4$\pm$23.2&7.63$\pm$7.07&57.2$\pm$21.8&3.70$\pm$2.71&73.9$\pm$12.9&3.76$\pm$2.35&1.91$\pm$1.94&72.9$\pm$11.1\\
DenseBiasNet \cite{he2019dpa}        &94.1$\pm$2.4&1.31$\pm$0.87&67.0$\pm$26.9&8.51$\pm$8.70&71.8$\pm$14.8&2.13$\pm$1.81&87.1$\pm$6.5&2.00$\pm$1.18&0.97$\pm$0.85&80.0$\pm$7.1\\
VFN (En) \cite{xia2018bridging}       &94.3$\pm$2.2&1.47$\pm$0.76&66.2$\pm$27.0&8.50$\pm$12.16&70.3$\pm$16.3&3.07$\pm$2.39&88.5$\pm$5.6&\textbf{1.36$\pm$0.64}&\textbf{0.53$\pm$0.46}&80.0$\pm$8.5\\
\hline
\textbf{Ours (En)}                      &\textbf{95.2$\pm$1.9}&\textbf{0.94$\pm$0.55}&\textbf{76.5$\pm$22.9}&\textbf{5.18$\pm$7.11}&\textbf{77.7$\pm$12.1}&\textbf{1.38$\pm$0.95}&\textbf{89.0$\pm$6.8}&1.66$\pm$0.89&0.69$\pm$0.66&\textbf{84.6}$\pm$\textbf{6.7}\\
\hline
\end{tabular}}
\label{tab1}
\end{table*}
\textbf{Quantitative evaluation.}
As shown in Tab.~\ref{tab1}, our proposed framework achieves the state-of-the-art performance. In our EnMcGAN, the artery achieves 89.0\% DSC, 1.66 MCD and 0.69 MSD which will strongly support the perfusion regions estimation and the arterial clamping position selection. The kidney gets 95.2\% DSC and 0.94 MSD, the tumor gets 76.5\% DSC and 5.18 MSD, and the vein gets 77.7\% DSC and 1.38 MSD which will provide operation guidance. The ensemble model, VFN, fuses the 2D information in different perspectives, achieving good performance on artery. However, similar to the V-Net, 3D U-Net, Res-U-Net, DenseBiasNet, it has poor performance on veins and tumors owing to the low contrast and the complex shapes.
\begin{figure*}[htb]
\centering
\includegraphics[width=11cm]{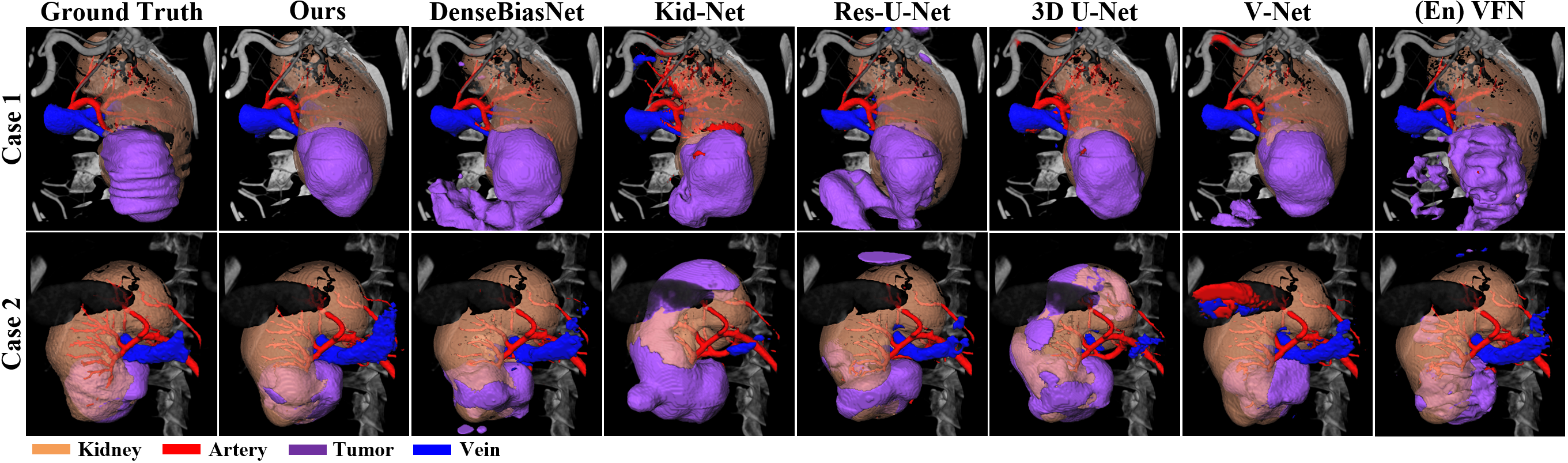}
\caption{Our proposed framework has powerful visual superiority which will provide visual guidance for surgery. Case 1 is a left kidney with angiomyolipoma and the case 2 is a right kidney with clear renal cell carcinomas.}
\label{fig3}
\end{figure*}

\textbf{Qualitative evaluation.}
As demonstrated in Fig.~\ref{fig3}, our proposed framework has great visual superiority which will provide visual guidance for surgery. Compared with the ground truth, our EnMcGAN enhances the contrast and utilizes the shape constraints from discriminator improving the integrity of tumors and the segmentation quality of veins. The Kid-Net and 3D U-Net have blur segmentation on small artery branches, and the DenseBiasNet, Res-U-Net, V-Net and VFN have serious under-segmentation because of the low contrast, complex shape and large anatomical variation in our CRS segmentation task.
\begin{figure}[htb]
\centering
\includegraphics[width=11cm]{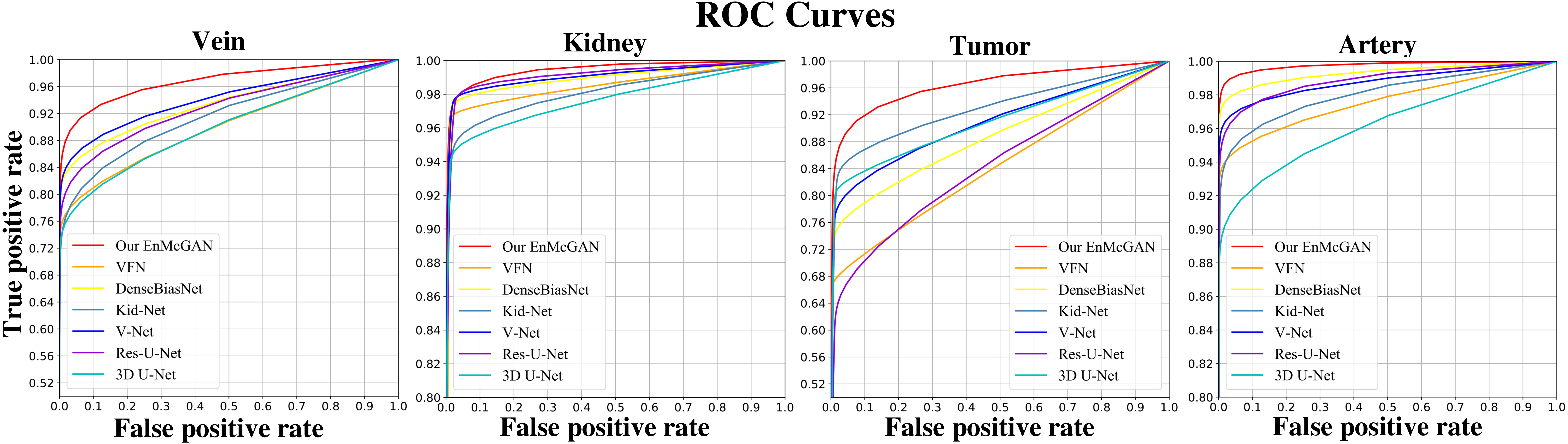}
\caption{The ROCs show that our EnMcGAN has higher segmentation accuracy than other comparison methods in our renal structures.}
\label{fig6}
\end{figure}

\textbf{ROC curves}
As shown in Fig.~\ref{fig6}, the ROC curve of our EnMcGAN covers the other comparison methods which means our proposed method has more powerful performance in each segmented renal structure. Due to the class imbalance in our task, the true positive rate will rise rapidly when plotting the ROC curve. For better demonstration, we show the ROC curves with a true positive rate between 0.8 and 1 in artery and kidney, and between 0.5 and 1 in vein and tumor.
\subsection{Ablation study shows improvements of the innovations}
\begin{table*}[htb]
\centering
\caption{The ablation study analyses the contributions of our innovations.}
\begin{tabular}{ccc|ccccc}
\hline
\multirow{2}{*}{McGAN}         &\multirow{2}{*}{MWC}    &\multirow{2}{*}{AWE}   &\multicolumn{5}{c}{DSC(\%)$\pm$std}\\
\cline{4-8}
&&                                    &Kidney       &Tumor           &Vein           &Artery            &AVG\\
\hline
            &            &            &94.1$\pm$2.4 &67.0$\pm$26.9   &71.8$\pm$14.8  &87.1$\pm$6.5      &80.0$\pm$7.1\\
\checkmark  &            &            &94.7$\pm$2.4 &75.0$\pm$22.7   &72.8$\pm$12.8  &87.6$\pm$8.1      &82.5$\pm$7.1\\
\checkmark  &\checkmark  &            &94.8$\pm$2.2 &75.8$\pm$21.1   &75.3$\pm$12.9  &\textbf{89.3$\pm$4.7}      &83.8$\pm$6.7\\
\checkmark  &\checkmark  &\checkmark  &\textbf{95.2$\pm$1.9} &\textbf{76.5$\pm$22.9}  &\textbf{77.7$\pm$12.1}   &89.0$\pm$6.8 &\textbf{84.6$\pm$6.7}\\
\hline
\end{tabular}
\label{tab:ablation}
\end{table*}
\label{Ablation}
The innovations in our framework brings significant enhancements. Our McGAN takes the shape constraints and improves 8.0\% DSC on tumor compared with the basic segmentation network (DenseBiasNet). When taking our MWC with the majority voting \cite{breiman1996bagging}, the vein and artery achieve 2.5\% and 1.7\% additional DSC improvement owing the fine-grained pattern in narrow windows. When adding our AWE module, the DSC of the tumor and vein are increased by 0.7\% and 2.4\% due to its dynamic ensemble weights directed at the input image. Totally, compared with the basic network, our EnMcGAN enhances the 1.1\%, 9.5\%, 5.9\% and 1.9\% DSC in kidney, tumor, vein and artery.
\subsection{Framework analysis}
\label{Framework}
\begin{figure}[htb]
\centering
\includegraphics[width=9cm]{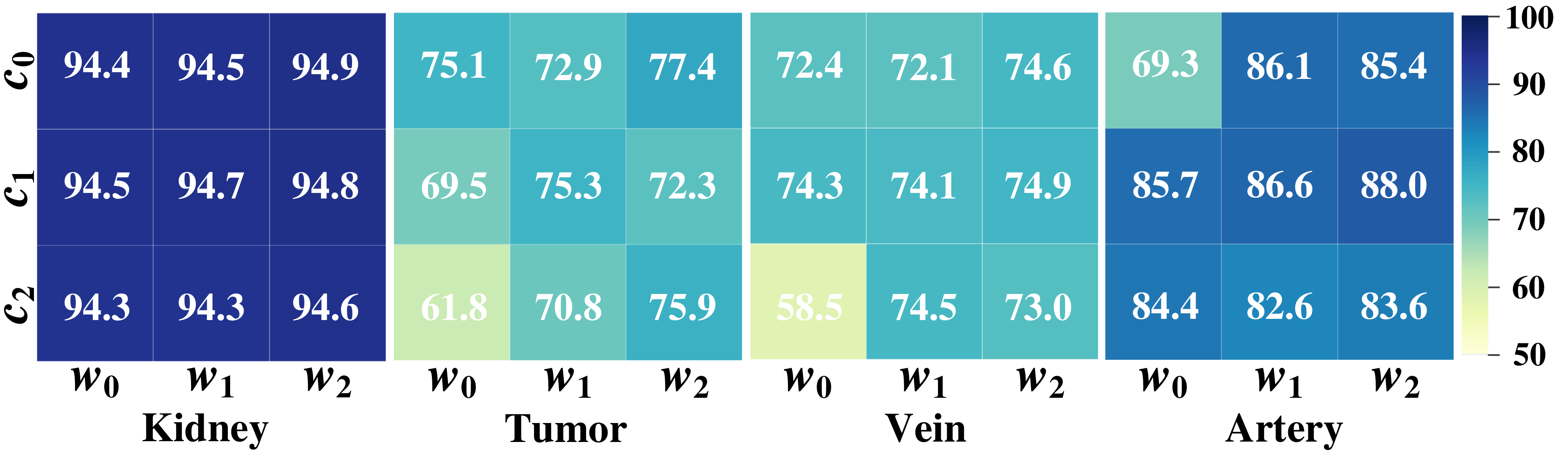}
\caption{The learners have segmentation superiorities in different narrow-window images. The heatmaps show the DSC of the learners on the renal structures.}
\label{fig4}
\end{figure}

\textbf{Segmentation learners analysis.}
As illustrated in Fig.~\ref{fig4}, two conclusions will be summarized: \textbf{1)} The learners have segmentation superiorities in different narrow-window images which have different salient regions. For example, when $c_{1}w_{2}$, the learner achieves significant performance on artery (88.0\%), but it has poor performance on tumor (72.3\%). The $c_{0}w_{2}$ has ordinary performance (85.4\%) on artery, but it achieves the highest DSC (77.4\%) on tumor. \textbf{2)} Fusing the segmentation superiorities of the learners will improve the whole segmentation quality which higher than the quality of each learner. Our EnMcGAN achieves 95.2\%, 76.5\%, 77.7\% and 89.0\% DSC on kidney, tumor, vein and artery which are higher than the best learners of renal structures (94.9\%, 77.4\%, 74.9\% and 88.0\%).

\begin{wrapfigure}{r}{0.4\linewidth}
\centering
\includegraphics[width=\linewidth]{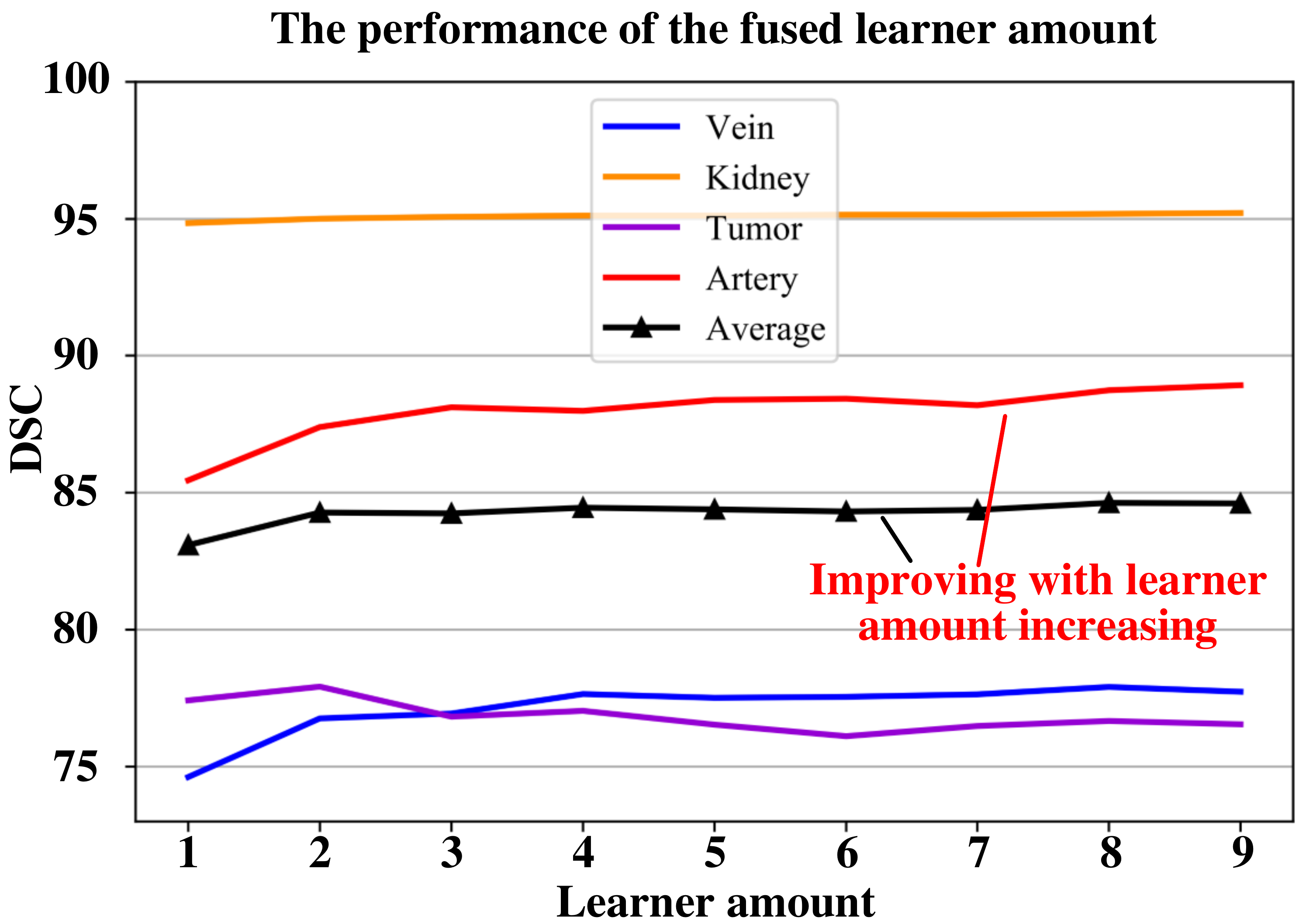}
\caption{With the amount of the fused learners increasing, the ensemble DSC will increase, and the structures have different sensitivity to the increasing of the learners.}
\label{fig5}
\end{wrapfigure}

\textbf{Amount of fused learners analysis.}
As shown in Fig.~\ref{fig5}, our with the amount of the fused learners trained in different sub-windows increasing, the ensemble accuracy will increase. We rank the trained segmentation learners in different sub-windows, and fuse them start from the best via our AWE strategy. It illustrate the characteristic in two aspect: 1) Overall, with the amount of the fused learners increasing, the average ensemble performance is increasing because more segmentation preferences are integrated into the model. 2) The renal structures has different sensitivity to the increasing of the learners. The performance of the kidney is almost unchanged because it has relatively large volume and wide gray range. The performance of tumor is increasing and decreasing, because it is sensitive to variation of window width and center. When the learner trained in the bad sub-window of tumor fused into the ensemble model, it will become the noise which will weaken the performance.
\section{Conclusion}
\label{Conclusion}
In this paper, we equips adversarial learning with ensemble segmentation models, and propose the EnMcGAN, the first 3D CRS segmentation model, for technical support of LPN. \textbf{1)}Our multi-window committee divides CTA image into narrow windows with different window centers and widths enhancing the contrast and making fine-grained pattern, and constructs the ensemble model based on these narrow windows fusing the segmentation superiorities on different covered distributions. \textbf{2)}Our multi-condition GAN utilizes the shape constraints of adversarial losses to encourage the segmented renal structures being consistent with their real shape, thus segmentation will tend to extract the features that matches the priori shape. \textbf{3)}Our adversarial weighted ensemble module uses the trained discriminator to score the quality of each structure from each learner for dynamic ensemble weights, enhancing the ensemble results. Extensive experiments with promising results reveal powerful 3D CRS segmentation performance and significance in renal cancer treatment. 

\subsubsection{Acknowledgements.}{This research was supported by the National Natural Science Foundation under grants (31571001, 61828101, 31800825), Southeast University-Nanjing Medical University Cooperative Research Project (2242019K3DN08) and Excellence Project Funds of Southeast University. We thank the Big Data Computing Center of Southeast University for providing the facility support on the numerical calculations in this paper.}
\bibliographystyle{splncs04}
\bibliography{mybib}
\end{document}